# Neutron Scattering "Halo" Observed in Highly Oriented Pyrolytic Graphite

Lilin He[1*], William Hamilton[2*], Tao Hong[3], Xin Tong[2], Lowell Crow[2], Katherine Bailey[2], Nidia Gallego[4]

[1] Biology and Soft Matter Division, Oak Ridge National Laboratory, Oak Ridge, Tennessee 37831, USA

[2] Instrument and Source Division, Oak Ridge National Laboratory, Oak Ridge, Tennessee 37831, USA

[3] Quantum Condensed Matter Division, Oak Ridge National Laboratory, Oak Ridge, Tennessee 37831, USA

[4] Materials Science & Technology Division, Oak Ridge National Laboratory, Oak Ridge, Tennessee 37831, USA

* Corresponding authors. Email: hel3@ornl.gov, hamiltonwa@ornl.gov

*This report has been authored by UT-Battelle, LLC under Contract No. DE-AC05-00OR22725 with the U.S. Department of Energy. The United States Government retains and the publisher, by accepting the article for publication, acknowledges that the United States Government retains a non-exclusive, paid-up, irrevocable, world-wide license to publish or reproduce the published form of this manuscript, or allow others to do so, for United States Government purposes. The Department of Energy will provide public access to these results of federally sponsored research*

**Abstract**

Highly oriented pyrolytic graphite (HOPG) has been widely used as monochromators, analyzers and filters at neutron and X-ray scattering facilities. In this Letter we report the first observation of an anomalous neutron "Halo" scattering of HOPG. The scattering projects a ring onto the detector with half cone angle of 12.4° that surprisingly persists to incident neutron wavelengths far beyond the Bragg cutoff for graphite (6.71Å). At longer wavelengths the ring is clearly a doublet with a splitting roughly proportional to wavelength. While the ring centers at the beam position if the beam is normal to the basal planes of HOPG, sample tilting results in the shift of the ring towards the same direction. The angle of ring shift is observed to be less than the sample tilts, which is wavelength dependent with longer wavelengths providing larger shifts. Additionally, upon tilting, the ring broadens and splits into a doublet at the low angle side with short wavelength neutrons whereas only subtle broadening is observed at longer wavelengths. The ring broadens and weakens with decreasing HOPG quality. We also notice that the intensity at the ring positions scales with the sample thickness. The ring vanishes as the sample is cooled down to 30 K, suggesting that the lattice dynamics of graphite is one of the factors that cause the scattering ring. A possible interpretation by combining inelastic scattering and "Yoneda" scattering is proposed.

## 1. Introduction

The atmospheric "rainbow"[1], "Halo"[2-3] and "Glory"[4] are well-known beautiful natural phenomena in meteorology. These phenomena, however, have different origins. The rainbow that often intensifies at ~42° is attributed to a refraction-reflection-refraction process of sunlight rays in water droplets[5-6]. "Halo", scientifically named parhelion, often near the sun or moon, arises from the refraction and reflection of light in small hexagonal ice crystals making up clouds in cold climates[6]. "Glory" has been interpreted as macroscopic light-tunneling effect by Nussenzveig[4, 6].

In materials science, neutron scattering techniques have proven to be powerful tools to investigate the microscopic structure and dynamics of materials[7-9]. These nondestructive techniques provide unique structural information on a wide variety of materials. At both neutron and synchrotron facilities, HOPG has been used as a monochromator, analyzer and filter [10-16]. This lamellar material is composed of many crystallites (mosaics) with honeycomb-like graphene layers well aligned along their *c*-axes whereas their *a*-axes are randomly oriented. The material is synthesized via heat treatment of carbon or by chemical vapor deposition at temperatures above 3000 °C under high pressures. Different mosaic spreads that determine HOPG qualities can be obtained by varying temperature, annealing time and deformation[10-11, 13, 17]. HOPG has also been used in imaging technology as a substrate and model electrode materials for batteries [18-21]. Despite being extensively used and studied for more than half a century, some new features of HOPG have never been reported until recently[22-25]. One particular example is that Ohmasa et al. observed radial-streak patterns when they studied HOPG using small-angle X-ray scattering (SAXS)[22]. The directions and appearance of these streaks changed with respect to the rotation of the samples, which was interpreted as double Bragg scattering.

During the course of an investigation into radiation damage in HOPG on the time-of-flight (TOF) Extended-Q Small-angle Neutron Scattering (EQ-SANS) [26] instrument at the Spallation Neutron Source (SNS) at ORNL, an unexpected ring feature of scattering was observed. This ring does not scale with wavelength and the ring profile broadens in longer wavelength frames. Subsequent measurements at General Purpose-SANS (GP-SANS)[27] and Bio-SANS[28] instruments (both are fixed-wavelength instruments) confirmed the observation. The ring feature was further investigated by a series of measurements including sample tilting, investigating other similar layered materials, and graphite samples with different mosaic spreads and different radiation damage. Inelastic neutron scattering measurements using a cold neutron triple-axis spectroscopy suggested that the ring arises from lattice dynamics coupled with "Yoneda" scattering instead of conventional diffraction. In the following sections, we report the details of these measurements and propose an explanation to the observed phenomena. To our best knowledge, this is the first report on "Halo" scattering phenomenon in the scattering process by a solid sample.

2. Experimental Section

*2.1 Materials*

Three HOPG samples with different grades were purchased from Union Carbide Corp. (TX, USA) and they arrive in strip shape. The ZYA HOPG (Mosaic ~0.4º FWHM) has dimension of 10mm × 70mm × 2mm, the ZYB HOPG (Mosaic ~0.8º FWHM) has dimension of 20mm × 100mm × 2mm and the ZYH HOPG (Mosaic ~1-2º FWHM) has dimension of 50mm × 40mm × 5mm. Five additional HOPG samples were supplied by Panasonic Corp. (Osaka, Japan), labeled as PGC×04, PGC×05, PGC×07, PGC×10 and PGC×20 with mosaic spread of $0.4º ≤θ<0.5º$, $0.5º ≤θ<0.6º$, $0.6º ≤θ<1.0º$, $1.0º ≤θ<2.0º$ and $2.0º ≤θ<3.0º$, respectively. These square samples have the same dimensions of 20mm × 20mm × 2mm. All samples were examined as received.

*2.2 Elastic and Inelastic Neutron Scattering Measurements*

SANS measurements were performed at all three SANS instruments at ORNL, including EQ-SANS, Bio-SANS and GP-SANS. The hexagonal lattice planes of the HOPG samples were normal to the incident beam direction except for the sample tilting measurements (Scheme. 1 left). In the measurements at EQ-SANS, two neutron wavelengths bands of 1.1−4.7 Å and 6.2−9.7 Å with a fixed sample-to-detector distance (SDD) at 2.0 m were used. At both Bio-SANS and GP-SANS, the detector was fixed at 2.2 m and wavelength was varied from 4.7 Å to 19 Å with a wavelength spread of ~ 15%.

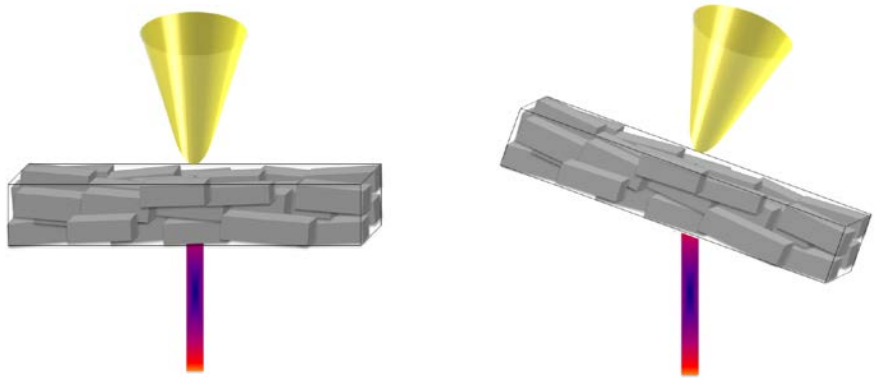

Scheme 1 Schematic view of the experimental setup, the cone indicates neutron trajectory of the ring feature. (left) with no sample tilting, the centers of the incident beam and the ring overlap; (right) After tilting the sample by 30°, the ring shifts towards the same direction as the sample tilts.

Inelastic neutron scattering measurements were carried out with the cold neutron triple-axis spectrometer (CTAX) at the high flux isotope reactor (HFIR), ORNL. The incident and final neutron wavelengths were selected by a PG (002) monochromator and analyzer, respectively. Contamination from higher-order reflections was removed by a cooled Be filter placed between the sample and the monochromator. Like the geometry in elastic scattering measurements, the sample was oriented in a way so that the crystallographic $c$ axis was parallel to the incident neutron beam. Inelastic neutron data were collected using the 3-axis mode with the incident neutron wavelength fixed at 4.0 Å, 4.5 Å and 5 Å, respectively. Elastic scattering data were collected without the use of the analyzer, with incident neutron wavelength fixed at 4.0 Å and 4.5 Å, respectively. Data was collected as a function of sample tilting angle and temperature.

## 3. Results and Discussion

*3.1 2D SANS patterns of HOPG using EQ-SANS*

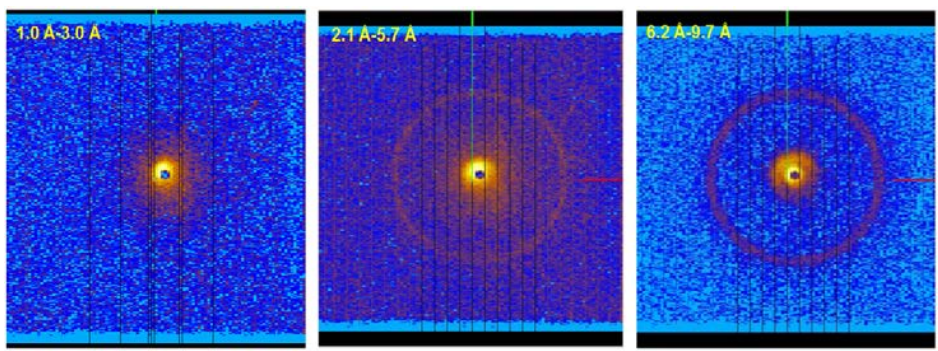

Figure 1 2D images of ZYA grade HOPG using EQ-SANS.

The 2D scattering pattern of ZYA grade HOPG sample collected on the Time-of-Flight SANS instrument of different wavelength bands are shown in Figure 1. At the wavelength band of 1.0-3.0 Å. No ring feature was observed. A weak but sharp ring started to show up for the wavelength frames of 2.1-4.7 Å and 6.2-9.7 Å (Figure 1 middle and right). The cone angle of this ring appears not to scale with wavelength. Instead, the angle remains a constant of about 12.4°, while the ring seemingly broadens with increasing wavelength (Figure 1 right).

*3.2 2D SANS patterns of HOPG vs Wavelength*

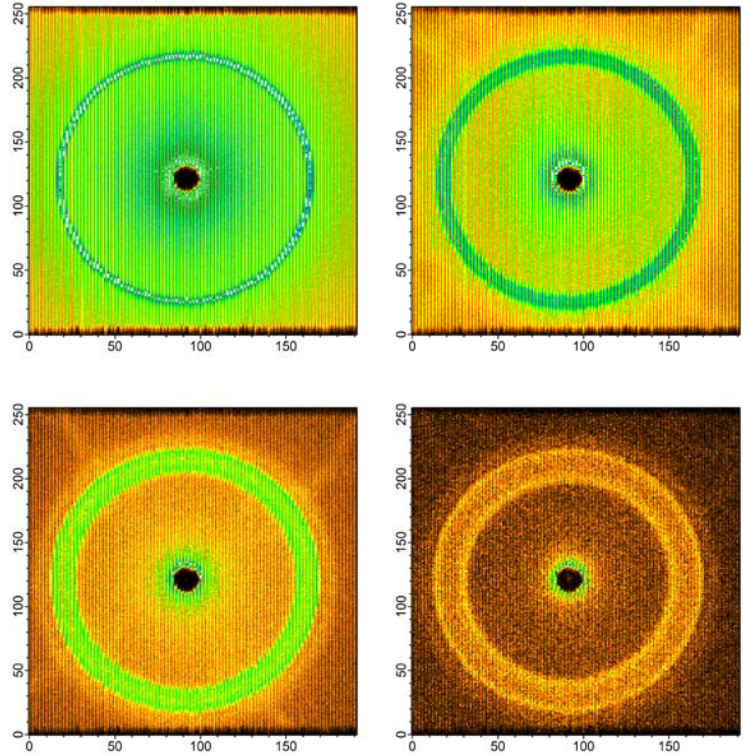

Figure 2 2D scattering images of PGC×04 HOPG as a function of wavelength using GP-SANS. The wavelength of incident neutron was 4.7 Å (upper left), 8 Å (upper right), 12 Å (bottom left) and 19 Å (bottom right), respectively. The wavelength spread for GP-SANS is ~15%.

This neutron scattering "Halo" effect was further verified using GP-SANS and Bio-SANS which have a monochromacy of ~15%. The same scattering ring appears in the SANS patterns of all HOPG samples made by Panasonic Corp.. The sample PGC×04 HOPG with the lowest mosaic spread shows ring broadening while the cone angle hardly changes with altering the wavelength from 4.7 Å to 19 Å (Figure 2), indicating the ring does not originate from Bragg diffraction. Additionally, the images collected using wavelength 8Å and 12Å show vague but noticeable radial streaks, which we believe correspond to the double Bragg scattering as reported by Ohmasa et al..[22] Line profiles reveal that the ring splits into a doublet with angular separation increasing with wavelength (Figure 3). The increase in wavelength also leads to the appearance of the secondary peak that corresponds to the fainter ring in the 2D images. While the location of this peak is seemingly independent of wavelength, the relative intensity of this peak to the primary one increases gradually as wavelength increases

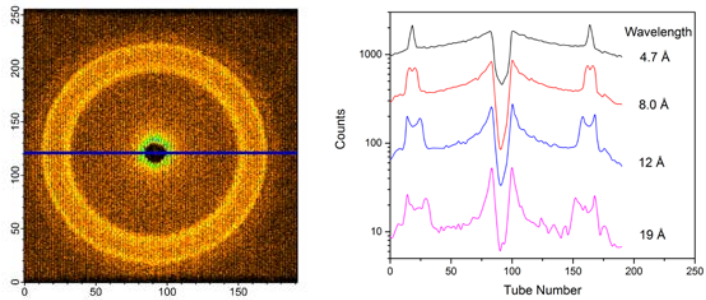

Figure 3 Line profiles crossing the ring at different wavelengths

*3.3 2D SANS patterns of HOPG, Mica and Germanium*

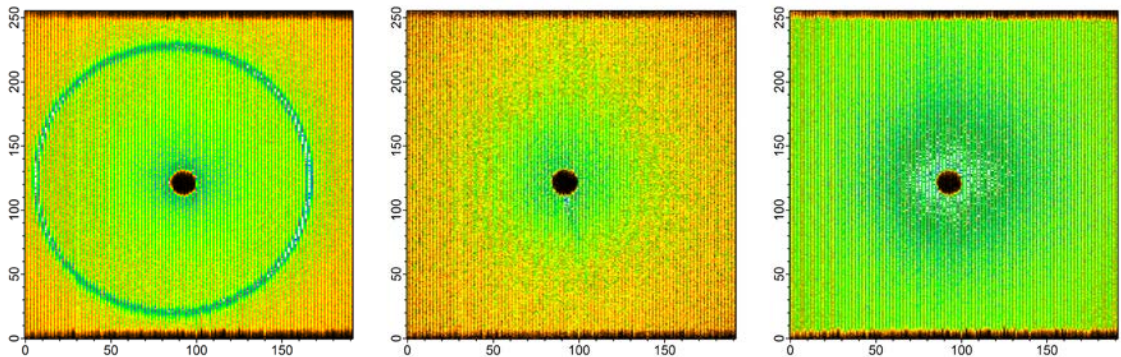

Figure 4 2D images of ZYA grade HOPG (left), Mica (middle) and single crystal Germanium (right).

Other crystal materials with similar layered structures as HOPG were also examined to further check the nature of the ring. Figure 4 displays SANS images of ZYA grade HOPG (left), Mica (middle) and single crystal Germanium (right) taken at 4.7 Å and 1.7 m SDD with zero detector translational offset. The mosaics of above materials are all highly ordered along *c* axis and disordered in the basal planes. Interestingly the ring was observed only for the HOPG sample. Hence, we conclude that the ring reflects unique structural and dynamic features of HOPG samples.

*3.4 Cone shifting angle vs sample tilting angle*

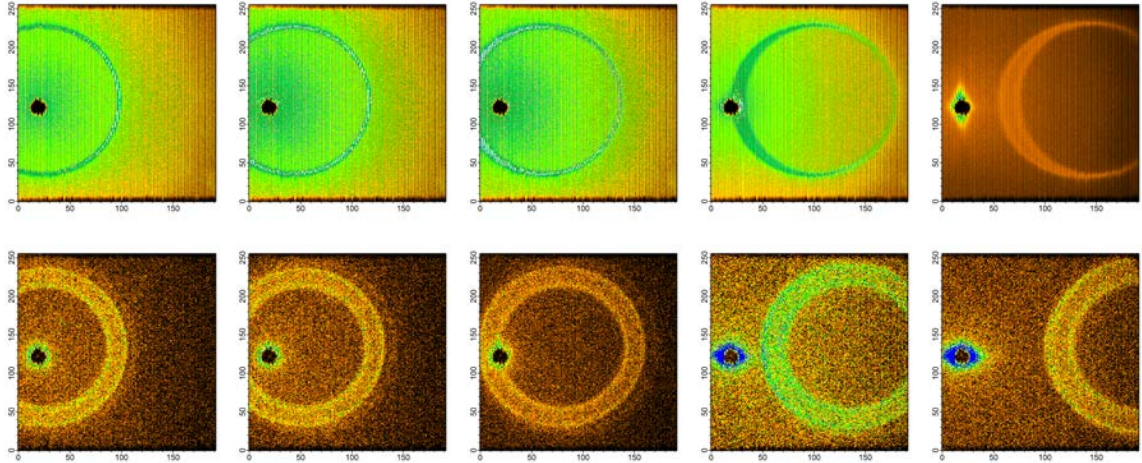

Figure 5 2D images of PGC×04 HOPG as a function of sample tilting angle (0°, 5°, 10°, 20° and 30° from left to right). The wavelength is 4.7 Å (upper row) and 19 Å (bottom row). A 40 cm detector translational offset was used for observing the cone shifting at large sample tilting angles.

The ring cone shifts in the same direction as the sample surface normal moves (scheme 1, right), which can be clearly seen in the 2D images from PGC×04 HOPG (Mosaic ~0.4° FWHM) measured at various tilting angles as shown in Figure 5. The upper row was measured with 4.7 Å neutrons whereas the bottom row was obtained using 19 Å neutrons. The cone angle shifts nearly linearly with the tilting angle of the sample (Figure 6). Interestingly the shift/tilt ratio is dependent on the wavelength and the values change from 0.61 at 4.7 Å, 0.73 at 12.0 Å to 0.89 at 19 Å. This ratio significantly deviates from the value of 2 which is expected according to the Bragg Law. The linearity between the wavelength and the ratio suggests that this ring effect could be used for wavelength calibration without the use of chopper system. For 4.7 Å neutrons, the images show that the ring broadens and retains its circular shape with increasing sample tilting angle. Moreover, the largest measured tilting angle (30 degrees) reveals that it is a doublet at low angle side. No significant changes, however, are observed with tilting the sample for 19 Å neutrons.

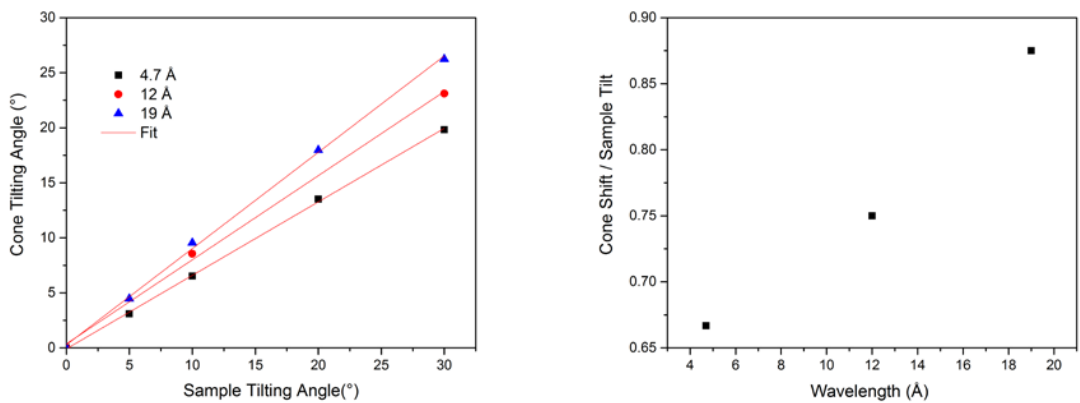

Figure 6 (a) Cone tilting angle *vs* sample tilting angle at different incident wavelengths; (b) The ratio of cone shift and sample tilting angle as a function of wavelength

*3.4 Ring broadening with sample quality (mosaic spread and radiation damage)*

*3.4.1 Mosaic spread*

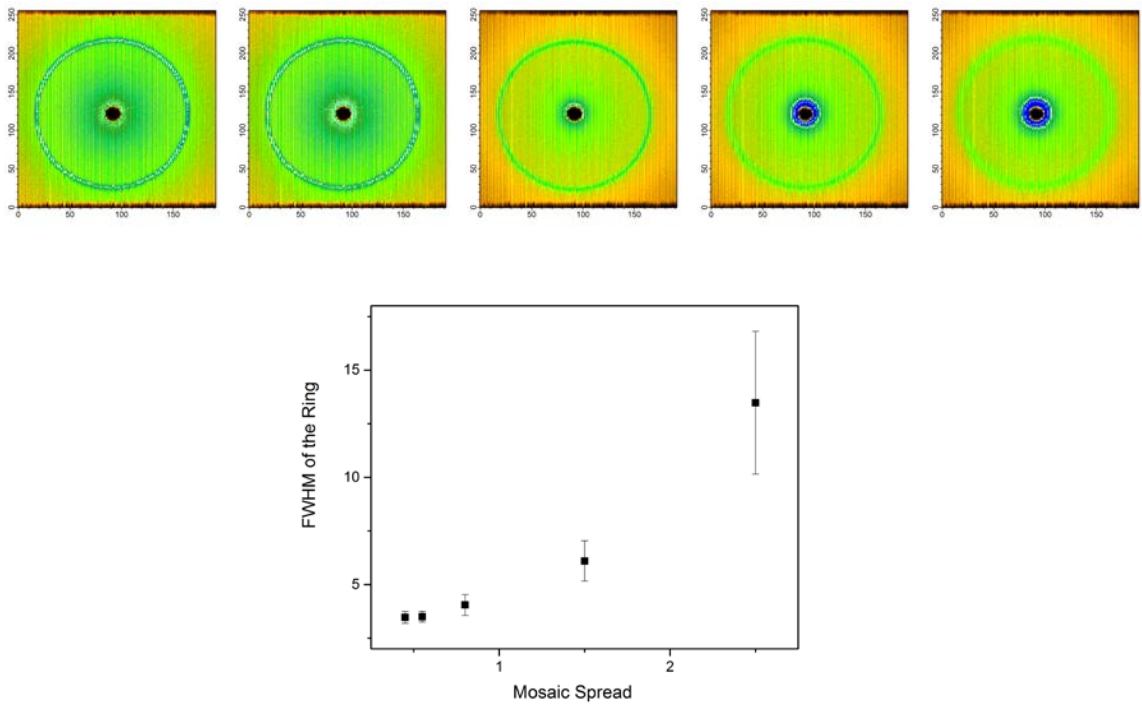

Figure 7 2D images of HOPG samples with different mosaic spreads (upper) and FWHM of the ring peak as a function of mosaic spread (bottom). The FWHM was obtained by fitting the peak using a Gaussian function.

As one would expect the ring "width" does broaden with decreasing HOPG quality (Figure 7). The full width at half maximum (FWHM) obtained by fitting the peak using a Gaussian function increases exponentially with the mosaic spread.

*3.4.2 Radiation damage effect*

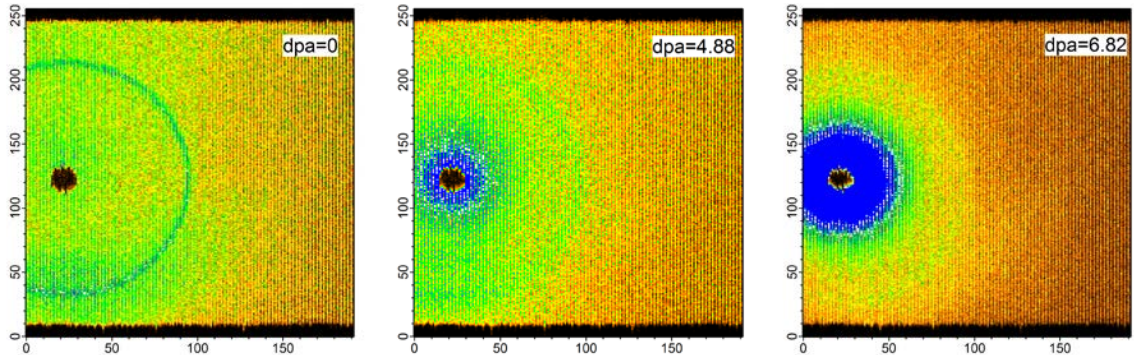

Figure 8 2D images of HOPG samples with different neutron irradiation doses (dpa)

The ring appears in all HOPG samples including the ones with no previous exposure to the beam. It seems, however, that the ring is weakened if the samples are damaged by radiation. Figure 8 presents 2D images from three samples that were exposed to different irradiation dose: 0, 4.88 and 6.82 displacement per atom (dpa) from left to right, respectively. The sharp ring broadens and fades away for the samples with higher irradiation levels, which is attributed to the structural disorder induced by the radiation in the crystal.

*3.5 Elastic and inelastic neutron scattering measurements using triple axis spectrometer*

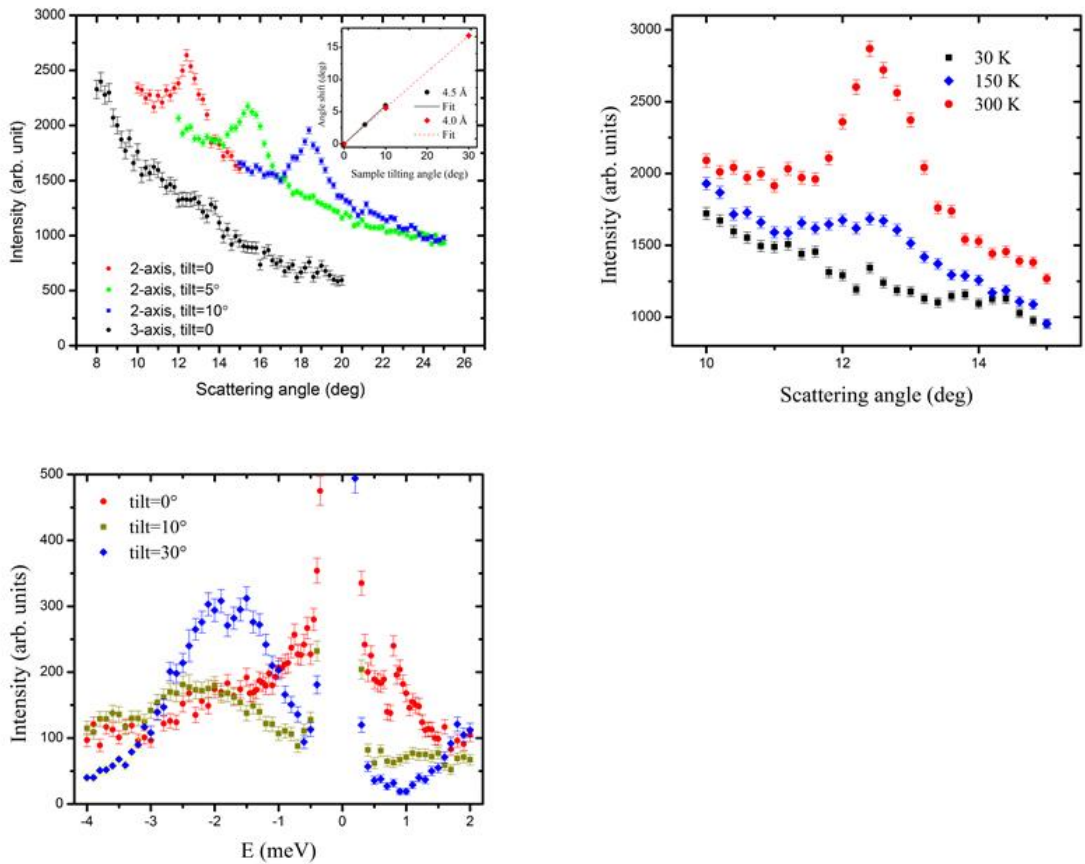

Figure 9 Inelastic neutron scattering measurements on a cold neutron triple-axis spectrometer. (a) 2-axis and 3-axis room-temperature measurements of the scattering angle scans at different tilt angles and different wavelengths on HOPG; the inset is the ring shift as a function of sample tilting for 4.0 Å and 4.5 Å, respectively; (b) the scattering angular distribution as a function of temperature for 4.0 Å using 2-axis mode; (c) the neutron scattering intensity *vs* energy transfer at the peak position on the high angle side at three tilting angles and room temperature for 4.0 Å.

To better understand the origin of the ring cone observed in SANS, we performed elastic and inelastic neutron scattering measurements on a cold-neutron triple-axis spectrometer (CTAX) at HFIR and results are shown in Fig. 9. Figure 9a represents the sample scattering angle scans at the incident neutron wavelength $\lambda_i$ ~4.5 Å with three-axis diffraction mode and 2-axis mode and different sample tilting angles. Data in 3-axis diffraction mode shows no sign of any peak. Subsequently the analyzer was moved away and data were collected in the two-axis mode, which in principle integrates over all low energies of scattered neutrons. Data at zero sample titling angle clearly show a peak at 12.4°. This peak shifts with the sample tilting. The observed sample tilting angle dependence of the peak position for two wavelengths, 4.0 Å and 4.5 Å, shown in the insert of Figure 9a, is in a good agreement with that from SANS. The above results suggest the origin of the observed ring cone at SANS is not static but dynamic. Further evidence is indicated in the follow-up inelastic neutron scattering measurements. Figures 9b shows the energy transfer spectrum at room temperature as a function of transferred energy at $\lambda_i$=4.0 Å, 4.5 Å and 5.0 Å, respectively, without tilting the sample. The observed excitations are attributed to the contributions from low energy vibration modes[29]. The peak intensity of 4.0 Å at room temperature diminishes with reducing temperature, which is displayed in Figure 9b. This provides unambiguous evidence that the lattice dynamics of HOPG should be one of the factors that cause this ring due to the fact that the thermal population factor of phonon decreases with decreasing temperature. Excitation energies of phonon modes are clearly altered by tilting the sample (Figure 9c).

## 4. Discussion

The "Halo" angle of 12.4° hardly changes with increasing wavelength from 4.7 Å to 19 Å (Fig. 2), which rules out pure diffraction interpretation due to the sample's internal structure. Neutron inelastic scattering measurements indicate unequivocally that the ring is related with lattice vibrations (Fig.9b). Although the ultimate answer of this anomalous "Halo" phenomenon is still unclear, here we present our hypothesis with a combination of both inelastic scattering and neutron reflection.

The neutrons are scattered weakly in all directions due to coherent inelastic scattering of HOPG, which provide the source for the subsequent scattering process. Phonons, the quantized atomic movements in a crystalline solid according to quantum mechanics, can absorb or emit energy when interacting with neutrons. In most solids, the meV energy of acoustic phonons matches the energy of neutrons used in inelastic neutron scattering measurements, allowing this technique for determining energy transfer during the scattering process[30-31]. The angular and energy distribution of inelastic neutrons are determined by the conservation laws [30].

$$k^2 = \left(\frac{2m}{h}\right)\omega_i(q)$$

where $k$ and $m$ are the wave vector and mass of exiting neutron respectively, $q$ is momentum transfer, $j$ is polarization index, $\omega$ is the energy of the absorbed phonon. For coherent one-phonon scattering by a single crystal, neutrons could be scattered in all directions and the outgoing energy in each direction has a finite number of discrete values.[30]

The total reflection of neutrons by specific planes in HOPG produces the "Halo" ring. The angle between the (101) plane or the ($\bar{1}$01) plane in the stacked graphene layers and $c$ axis of the crystallites is ±12.3° as shown in the left image of Fig. 10. This angle matches well with the "Halo" angle within the instrument uncertainty. The "Yoneda" scattering, in which neutrons are reflected by a rough surface at an angle slightly greater than the critical angle, will produce a peak [32-35]. Sinha et al. has ascribed the physical origin of "Yeneda" scattering to the fact that the electric field at the surface reaches a maximum once the incident angle is close to the critical angle [36]. Rotational symmetry of such a peak around $c$ axis in HOPG eventually gives rise to a perfect ring we observe. The peak position, which is at ~ 12.4° in this case with respect to the beam direction, is nearly independent on incident wavelength. However, increasing wavelengths causes the ring broadening due to the increase in critical angle. One can calculate the critical angle via: $\theta_c = \lambda\sqrt{\beta/\pi}$ where $\beta$, the scattering length density of HOPG, equals $7.55\times10^{-6}$ Å$^{-2}$. The calculated critical angles for 4.7 Å, 8 Å, 12 Å and 19 Å are 0.42°, 0.71°, 1.07° and 1.69° respectively. The observed doublet feature (Fig. 3) at long wavelengths can be explained as the "Yoneda" scattering by two opposite faces of the (101) plane or the ($\bar{1}$01) plane (Figure 10 right). Because the separated rings cannot be resolved for 4.7 Å due to the low instrument spatial resolution, a singlet is observed. Furthermore, when cooling the sample, the number of inelastically scattered neutrons drops dramatically, thus, the ring disappears.

When the sample is tilted, the ring cone shifts nearly linearly with the tilting angle as presented in Figure 6a. The shift/tilt ratio is sensitive to the wavelength of incident neutrons. More interestingly, 4.7 Å neutron exhibits different ring patterns than larger wavelength neutrons (Fig. 5): the ring at low angle side broadens and splits into a doublet while the ring at high angle side fades. Furthermore, the doublet separation increases with increasing the tilting angle. The separation reaches ~2.4° at the 30° tilting angle. It is also noticeable that the broadening of the peak takes place for 19 Å neutrons as tilting the sample, although it is not significant. In order to understand the origin of this separation, we have tried to measure the energy transfer as a function of the tilting angle using 4.0 Å neutrons. Unfortunately, the inelastic scattering measurement on the low angle side where the separation occurs failed because it was too close to the direct beam. The data displayed in Figure 9d was collected on the high angle side. Also, the energy-loss branch of the spectra was just partially attained due to the resolution limit of the instrument. One observes an increasing signal of energy gain with increasing tilting angle as shown in Fig. 9c. For the 30° tilting angle, the energy gain peaks at ~ 1.8 meV. The energy loss peak, on the other hand, displays a maximum at ~2 meV within the measured energy transfer range. The energy gain or loss of neutrons alters the incident wavelength for the subsequent "Yoneda" scattering, which may eventually result in the doublet ring. For 4.7 Å neutrons, our calculations show that the critical angles will change from 0.42° to 0.33° if gaining 2 meV and to 0.61° if losing 2 meV, respectively. Apparently the energy transfer of ~2 meV is not sufficient to account for the 2.4° doublet separation. Presumably stronger vibration mode that induces much larger energy transfer is responsible for it, which is out of our measuring range. Further investigation is required to verify the hypothesis.

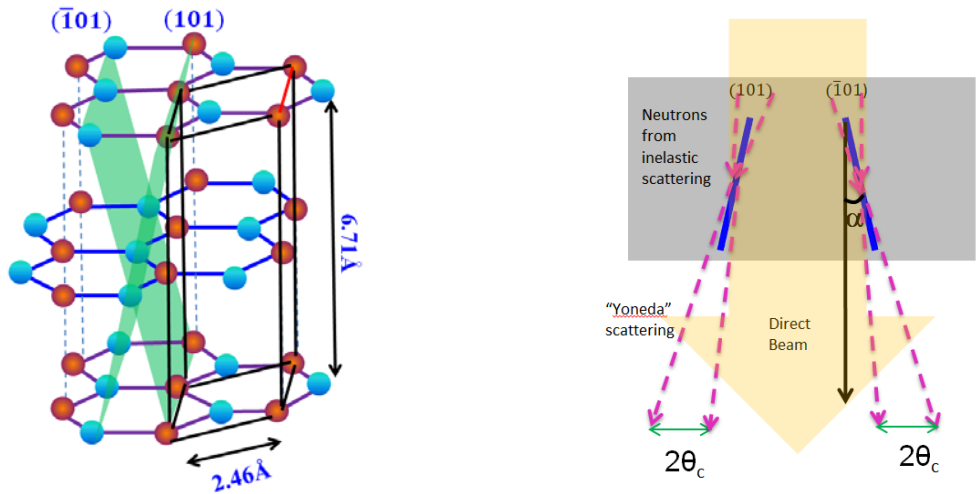

Figure 10 (a) Structure of stacked graphene layers with lattice constants $a_0$ =2.46 Å and $c_0$=6.71 Å. (b) Schematic picture of "Yoneda" scattering by the planes of (101) and ($\bar{1}$01) to give rise to the doublet ring.

In the above interpretations we have been concerned with one-phonon coherent scattering only. It is much more complicated for multiple scattering processes although it is very unlikely for HOPG samples less than 2mm thick. Nevertheless, one cannot distinguish single scattering from all other processes via experiments. Additionally, it is still not clear that the ring is absent for neutrons less than 3 Å. The ultimate explanation of this "Halo" phenomenon may require a wider range of energy scan and first principle calculations of phonon dispersion of HOPG.

## Summary

In summary, we observed an anomalous scattering ring in the SANS pattern of HOPG. The findings are summarized as follow:
1. The ring cone angle is nearly independent of the incident wavelength;
2. No ring was observed for the wavelength less than ~ 3Å;
3. The scattering process involves inelastic scattering;
4. The ring intensity is weak, but sharp for short wavelength, it broadens with increasing wavelength, the ring apparently is a doublet for longer wavelengths;
5. The ring shifts with sample tilts, and the cone tilting angle follows nearly linearly with sample tilting angle depending on the wavelength; The sharp ring of short wavelength (4.7 Å) broadens and splits into a doublets on the side of direct beam; It diminishes on the other side at sufficiently large tilting angles;
6. The ring broadens and weakens with decreasing crystallite quality;
7. The peak intensity scales with the sample thickness.

All the evidences indicate that the ring might be caused by the tandem effects of lattice vibration and total reflection in graphite. However, the intrinsic mechanism that leads to this scattering ring is not definitive at present. We expect ongoing wider range of energy scan and first principle calculations of phonon dispersion of HOPG with a focus on low frequency modes to provide more clues for completely solving this puzzle. This study also indicates that the lattice vibration of solids especially crystals can contribute significantly to the scattering signals especially in the small angle range, and such contribution could be temperature and angle dependent. Knowledge of the existence of "Halo" rings are important and care should be taken in data analysis as they could be labeled as diffraction peaks especially in 1D data. Beyond possible complication this

"Halo" effect brings to data analysis, it may emerge as a promising tool for beam and sample alignment for SANS instruments. The effect reported in this Letter may have considerable implications in materials and neutron optical subfields.

**Acknowledgements**

This research used resources at the Spallation Neutron Source and the High Flux Isotope Reactor, DOE Office of Science User Facilities operated by the Oak Ridge National Laboratory. Authors thank Panasonic Corp. for providing HOPG samples with different mosaic spreads. We also thank Drs. Changwoo Do and Shuo Qian for their help on the measurements at EQ-SANS and Bio-SANS. Drs. Volker Urban, Stephen Nagler, Huibo Cao, Yaohua Liu, Feng Ye, and Songxue Chi at ORNL, Roger Pynn at Indiana University Bloomington, and Chuck Majkrzak at NIST are acknowledged for their insightful discussions.